\documentclass[aps, prl, twocolumn, superscriptaddress, floatfix]{revtex4-2}
\usepackage{graphicx}
\usepackage{amsmath}
\usepackage{ulem}
\usepackage{amssymb}
\usepackage{tikz}
\usepackage{marvosym}
\usepackage{braket}
\usepackage{float}
\usepackage{pgfplots}
\usepackage{mathrsfs}
\usepackage[version=4]{mhchem}
\pgfplotsset{compat=1.17}
\usepackage{subcaption}
\usetikzlibrary{shapes.geometric}
\usetikzlibrary{shadows,patterns,perspective}
\usetikzlibrary {decorations,decorations.text}
\usetikzlibrary{decorations.pathreplacing}
\usepackage{float}
\usepackage{hyperref}
\usepackage{xcolor}

\definecolor{linkcolor}{RGB}{0, 0, 255}      % Blue color for links
\definecolor{citecolor}{RGB}{0, 128, 0}     % Green color for citations
\definecolor{urlcolor}{RGB}{255, 0, 0}       % Red color for URLs

\hypersetup{
    colorlinks=true,
    linkcolor=linkcolor,
    citecolor=citecolor,
    urlcolor=urlcolor,
    linktoc=all,  % Set links in the table of contents to be colored as well
    pdfborder={0 0 0}  % Removes the box around links
}

\DeclareCaptionJustification{justified}{\leftskip=0pt \rightskip=0pt \parfillskip=0pt plus 1fil}

\begin{document}
\definecolor{dy}{rgb}{0.9,0.9,0.4}
\definecolor{dr}{rgb}{0.95,0.65,0.55}
\definecolor{db}{rgb}{0.5,0.8,0.9}
\definecolor{dg}{rgb}{0.2,0.9,0.6}
\definecolor{BrickRed}{rgb}{0.8,0.3,0.3}
\definecolor{Navy}{rgb}{0.2,0.2,0.6}
\definecolor{DarkGreen}{rgb}{0.1,0.4,0.1}

\title{Two-Channel Kondo behavior in the quantum XX chain with a boundary defect}

\author{Yicheng Tang}
\email{tang.yicheng@rutgers.edu}
\affiliation{Department of Physics and Astronomy, Center for Material Theory, Rutgers University,
Piscataway,  New Jersey, 08854, United States of America}
\author{Pradip Kattel}

\affiliation{Department of Physics and Astronomy, Center for Material Theory, Rutgers University,
Piscataway, New Jersey, 08854, United States of America}
\author{J. H. Pixley}
\affiliation{Department of Physics and Astronomy, Center for Material Theory, Rutgers University,
Piscataway,  New Jersey, 08854, United States of America}
\affiliation{Center for Computational Quantum Physics, Flatiron Institute, 162 5th Avenue, New York, NY 10010}
\author{Natan Andrei}
\affiliation{Department of Physics and Astronomy, Center for Material Theory, Rutgers University,
Piscataway,  New Jersey, 08854, United States of America}

\begin{abstract}
We demonstrate that a boundary defect in the single spin-$\frac{1}{2}$ quantum $XX$ chain exhibits two-channel Kondo physics. Due to the presence of the defect, the edge spin fractionalizes into two Majorana fermions, out of which one decouples, and one is overscreened by the free fermion in bulk, leading to non-trivial boundary behavior characteristic of the two-channel Kondo model. When the boundary-to-bulk coupling ratio exceeds a critical value of $\sqrt{2}$, a massive boundary-bound mode is exponentially localized near the impurity site for strong impurity coupling. This leads to unusual behavior in physical quantities, such as the $g$-function not being monotonic. We compute the $g-$function of the impurity from thermodynamic and entanglement entropy calculations and show that it takes a non-integer value of $\sqrt{2}$ just as in the two-channel Kondo problem.
\end{abstract}

\maketitle
Exotic quasiparticles exist in a plethora of many-body interacting systems \cite{piton,fractonrm,wrinklon,polaron,plasmon,exciton,axion}. Some quasiparticles bear fractional charges as in fractional quantum Hall systems \cite{stormer1999fractional,moore1991nonabelions}, Majorana anyons in the Kitaev chain \cite{kitaev2001unpaired} and the two-channel Kondo problem \cite{BAandrei1984solution,BAtsvelick1985exact, CFTludwig1994exact,CFTaffleck1993exact,emery1992mapping,mitchell2012universal}, the Fibonacci anyons in the topological defects of the Potts model \cite{chang2019topological} and the three-channel Kondo problem\cite{cftla,CFTaffleck1995conformal,PhysRevLett.113.076401}, or more exotic non-Abelian anyons in   multichannel Kondo problem \cite{lopes2020anyons,PhysRevLett.128.146803}.

The multichannel Kondo problem is one of the most studied models in condensed matter systems, which can be solved by a variety of non-perturbative methods like Wilson's numerical renormalization group \cite{NRwilson1975renormalization,NRaffleck1992relevance,NRtoth2009numerical,NRbulla2008numerical}, Bethe Ansatz \cite{BAandrei1984solution,BAtsvelick1985exact,BAjerez1998solution}, boundary conformal field theory (bCFT) \cite{CFTaffleck1993exact,CFTludwig1994exact,CFTaffleck1995conformal}, and a careful large-S analysis \cite{krishnan2024kondo}. It is well known that in the original multichannel Kondo problem, where multiple flavors (channels) of electrons interact with a localized spin, the charge and flavor degrees of freedom do not couple directly to the impurity—only the spin of the conduction channels does \cite{ANDREI1984108, CFTaffleck1995conformal}. More recently, similar decoupling has been observed in a variety of other settings, including charge Kondo systems~\cite{iftikhar2015two,iftikhar2018tunable,pouse2023quantum,karki2023z} and spin defects in spin chains \cite{ANDREI1984108, kattel2023kondo,kattel2024kondo,schlottmann1991impurity,sacramento1993thermodynamics,liu1997low,laflorencie2008kondo,wang1997exact,frahm1997open,giuliano2018kondo}. 
In this work, we show that the spin-$\frac{1}{2}$ XX quantum spin chain with a boundary defect exhibits two-channel Kondo physics. We show that the chain hosts decoupled Majoranas at its boundary \cite{emery1992mapping,lopes2020anyons,coleman1995pedestrian}, exhibiting non-Fermi liquid behavior and the expected zero-temperature entropy  $k_B\ln\sqrt{2}$ in the thermodynamic limit just as in the two-channel Kondo problem. 
% Moreover, upon Jordon-Wigner transformation, the fermionic Hamiltonian with an odd number of bulk sites shows double degeneracy throughout its entanglement spectra~\cite{calabrese2008entanglement} for entanglement cuts between even and odd sites in the bulk, just as in the topological spin chain. However, the entanglement spectrum has no perfect double degeneracy if the entanglement cuts are made between the odd and even sites in bulk.
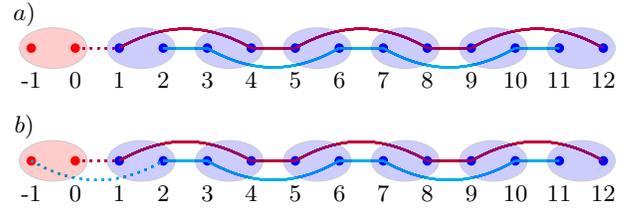
\begin{figure}
    \centering
\begin{tikzpicture}
   \begin{scope}[scale=0.9]
        \def\n{14} 
    \def\spacing{0.65}
    \foreach \i in {0,...,13} {
        \ifnum\i<2
            \fill[red] (\i*\spacing, 0) circle (2pt);
        \else
            \fill[blue] (\i*\spacing, 0) circle (2pt);
        \fi
        \draw[thick,purple, dotted] (1*\spacing, 0) -- (2*\spacing, 0);
        \draw[thick, bend left=30,purple] (2*\spacing, 0) to (5*\spacing, 0);
        \draw[thick,purple] (5*\spacing, 0) -- (6*\spacing, 0);
            \draw[thick, bend left=30,purple] (6*\spacing, 0) to (9*\spacing, 0);
        \draw[thick,purple] (9*\spacing, 0) -- (10*\spacing, 0);
            \draw[thick, bend left=30,purple] (10*\spacing, 0) to (13*\spacing, 0);
        \draw[thick,cyan] (3*\spacing, 0) -- (4*\spacing, 0);

    \draw[thick, bend right=30,cyan] (4*\spacing, 0) to (7*\spacing, 0);

    \draw[thick,cyan] (7*\spacing, 0) -- (8*\spacing, 0);

    \draw[thick, bend right=30,cyan] (8*\spacing, 0) to (11*\spacing, 0);

\draw[thick,cyan] (11*\spacing, 0) -- (12*\spacing, 0);
        \node[below, yshift=-6pt] at (\i*\spacing, 0) {\the\numexpr-1+\i\relax};
    }
    \draw[fill=red, opacity=0.2] (0.5*\spacing, 0) ellipse (0.5 and 0.3);
    \draw[fill=blue, opacity=0.2] (2.5*\spacing, 0) ellipse (0.5 and 0.3);
    \draw[fill=blue, opacity=0.2] (4.5*\spacing, 0) ellipse (0.5 and 0.3);
    \draw[fill=blue, opacity=0.2] (6.5*\spacing, 0) ellipse (0.5 and 0.3);
    \draw[fill=blue, opacity=0.2] (8.5*\spacing, 0) ellipse (0.5 and 0.3);
    \draw[fill=blue, opacity=0.2] (10.5*\spacing, 0) ellipse (0.5 and 0.3);
    \draw[fill=blue, opacity=0.2] (12.5*\spacing, 0) ellipse (0.5 and 0.3);
        \node at (-0.1,0.5) {$a)$};

   \end{scope}

    \begin{scope}[yshift=-1.5cm,scale=0.9]
        
    \def\n{14} 
    \def\spacing{0.65}
    \foreach \i in {0,...,13} {
        \ifnum\i<2
            \fill[red] (\i*\spacing, 0) circle (2pt);
        \else
            \fill[blue] (\i*\spacing, 0) circle (2pt);
        \fi
        \draw[thick,purple, dotted] (1*\spacing, 0) -- (2*\spacing, 0);
        \draw[thick, bend left=30,purple] (2*\spacing, 0) to (5*\spacing, 0);
        \draw[thick,purple] (5*\spacing, 0) -- (6*\spacing, 0);
            \draw[thick, bend left=30,purple] (6*\spacing, 0) to (9*\spacing, 0);
        \draw[thick,purple] (9*\spacing, 0) -- (10*\spacing, 0);
            \draw[thick, bend left=30,purple] (10*\spacing, 0) to (13*\spacing, 0);
       
        \draw[thick, bend right=30,cyan, dotted] (0*\spacing, 0) to (3*\spacing, 0);
        \draw[thick,cyan] (3*\spacing, 0) -- (4*\spacing, 0);

    \draw[thick, bend right=30,cyan] (4*\spacing, 0) to (7*\spacing, 0);

    \draw[thick,cyan] (7*\spacing, 0) -- (8*\spacing, 0);

    \draw[thick, bend right=30,cyan] (8*\spacing, 0) to (11*\spacing, 0);

\draw[thick,cyan] (11*\spacing, 0) -- (12*\spacing, 0);
        \node[below, yshift=-6pt] at (\i*\spacing, 0) {\the\numexpr-1+\i\relax};
    }
    \draw[fill=red, opacity=0.2] (0.5*\spacing, 0) ellipse (0.5 and 0.3);
    \draw[fill=blue, opacity=0.2] (2.5*\spacing, 0) ellipse (0.5 and 0.3);
    \draw[fill=blue, opacity=0.2] (4.5*\spacing, 0) ellipse (0.5 and 0.3);
    \draw[fill=blue, opacity=0.2] (6.5*\spacing, 0) ellipse (0.5 and 0.3);
    \draw[fill=blue, opacity=0.2] (8.5*\spacing, 0) ellipse (0.5 and 0.3);
    \draw[fill=blue, opacity=0.2] (10.5*\spacing, 0) ellipse (0.5 and 0.3);
    \draw[fill=blue, opacity=0.2] (12.5*\spacing, 0) ellipse (0.5 and 0.3);
    \node at (-0.1,0.5) {$b)$};
    \end{scope}
\end{tikzpicture}
    \caption{{\bf Pictorial representation of the models considered in this work.} a) Hamiltonian $H_s$ in Eq.\eqref{MainH} with a decoupled Majorana Fermion with two free Majorana Fermion chains and b) Hamiltonian $H_s+J\sigma^y_0\sigma^y_1$ %Eq.\eqref{fig:pictorialHam1ch} 
    with two free Majorana Fermion chains, both with one impurity at the left end. The dashed line presents the impurity coupling $J$, whereas the solid lines represent the bulk coupling, which is set to be unity. The red oval represents the impurity with two Majorana labeled -1 and 0, and the blue ovals represent the bulk sites where the two blue circles in each oval represent the Majorana fermions.}
    \label{fig:pictorialHam}
\end{figure}

The following Hamiltonian describes the model under consideration
\begin{equation}
    H_s = \sum_{i=1}^{N-1} (\sigma^x_i\sigma^x_{i+1} + \sigma^y_i\sigma^y_{i+1}) +J\sigma^x_{0}\sigma^x_1,
    \label{modelham}
\end{equation}
where $i=1\cdots N$  label the bulk sites and $\vec\sigma_0$ are the Pauli matrices acting on the space of a single impurity located at the left edge of the chain, coupled to the first site of the chain via $\sigma_0^x\sigma_1^x$ interaction term and $J$ is the impurity coupling strength. Its value, we shall show,  determines two distinct impurity phases: when $J<\sqrt{2}$, the impurity spin is overscreened by the Kondo cloud and the low energy physics is described by an overscreened  Kondo conformal boundary fixed point, whereas when $J>\sqrt{2}$, the impurity is overscreened by a single particle bound mode exponentially localized at the edge of the chain. This phase does not admit a boundary CFT (bCFT) description, as the emergence of a massive boundary mode breaks conformal invariance, rendering the assumptions underlying the $g$-theorem inapplicable~\cite{friedan2004boundary,harper2024g}. Consequently, the $g$-function need not be monotonic in this regime.

In a recent paper \cite{kattel2024kondo}, we demonstrated that the Hamiltonian given by Eq. \eqref{modelham}, when coupled to an additional impurity interaction of the form $J \sigma^y_{0}\sigma^y_1$, exhibits single-channel Kondo behavior, as evidenced by the calculation of various thermodynamic quantities. In this work, we briefly outline the differences between the two models and provide a pictorial explanation for why the previous model yields single-channel behavior, whereas the current model exhibits two-channel Kondo behavior. It is worth noting that both models can be mapped to non-interacting Fermions via the Jordan-Wigner transformation as shown in Fig.~\ref{fig:pictorialHam}.

The nontrivial effect of the impurity becomes more apparent if the Hamiltonian is written in Fermionic variables using the Jordan-Wigner transformation, where we could apply the bCFT-inspired ideas to study the effect of the impurity as a boundary condition on the bulk fermion.  
% The Hamiltonian  in the Fermionic variables reads
% \begin{align}
%     H&=\sum_{j=1}^{N-1} c^\dagger_{j+1}c_j+c^\dagger_j c_{j+1}+ J\left(c_{1}c_0+c_{0}^\dagger c_1^\dagger + c_{0}^\dagger c_1+c^\dagger_1 c_{0} \right)
%     \label{hamFermionic}
% \end{align}

% Here, the bulk is a free fermion chain, and in the boundary, an impurity interacts with the fermion in the first site. To understand the effect of the boundary interaction in the system, performing another transformation and writing the Hamiltonian in Majorana fermion variables is more convenient.

Defining Majorana fermions as
\begin{equation}
    \gamma_{2l-1} = \prod_{m<l}\sigma^z_m \sigma^x_l \quad \gamma_{2l} = \prod_{m<l}\sigma^z_m \sigma^y_l \quad\quad l=0,1,\cdots,2N
\end{equation}
which satisfies $\gamma^2=1$ and $\{\gamma_l,\gamma_m\}=2\delta_{lm}$. The Hamiltonian reads
\begin{equation}
\label{MainH}
    H_m = \sum_{l=1}^{N-1} i(\gamma_{2l}\gamma_{2l+1}-\gamma_{2l-1}\gamma_{2l+2})+iJ\gamma_0\gamma_1,
\end{equation}
where we notice that $ \gamma_{-1} $ decouples~\cite{zvyagin2013possibility}, and the Hamiltonian splits into two independent Majorana chains in the bulk~\cite{perk1977time}. The Majorana language allows us to understand the role of boundary interaction more clearly. As shown in Fig.~\ref{fig:pictorialHam}, the bulk can be written as two decoupled free Majorana chains, and similarly, the impurity can be written as two Majorana fermions. Out of the two impurity Majorana fermions, one decouples, and another is coupled to one of the Majorana chains. 

The model captures the key characteristics of the two-channel Kondo problem. Specifically, a decoupled Majorana mode emerges, with only one-half of the impurity interacting with the fermionic bath. Moreover, only one of the Majorana chains undergoes a phase shift, while the other one remains unaffected. This is similar to %as in 
the two-channel Kondo model 
%case 
with %full 
spin, charge, and flavor (with $SO(3)\times SO(5)$ symmetry represented by eight different Majorana fields) degrees of freedom where 3 of the Majorana chains undergo a phase shift, whereas the remaining five do not \cite{ emery1992mapping, zhang1999majorana}. More concretely, the quasimomenta $k_{j,1}$  for the Majorana fermion chain that is not coupled to the impurity  depicted in cyan in Fig.~\ref{fig:pictorialHam} in the picture are given by
\begin{equation}
    k_{j,1} = \frac{\pi j}{N}\quad j=1,2,...,\frac{N}{2},
    \label{k1}
\end{equation}
when free boundary conditions are imposed, whereas the chain coupled to the impurity (shown with purple lines in Fig.~\ref{fig:pictorialHam}) has a spectrum with momentum-dependent phase shift due to the presence of the impurity. Its quasimomenta $k_{j,2}<\frac{\pi}{2}$ satisfy \cite{kattel2024kondo}
\begin{equation}
  e^{-2ik_{j,2}(N+1)}=A_B(k)=\frac{(1-J^2)+e^{2 i k_{j,2}}}{1 + (1-J^2)e^{2 i k_{j,2}}},
  \label{quantcond}
\end{equation}
with $A_B(k)$ being the boundary scattering amplitude.
Note that all momenta $k_{j,2}$ given by Eq.\eqref{quantcond} are real when $J<\sqrt{2}$, however beyond it there appears a complex solution of the form $k_{j,2}=\frac {1} {2} i\ln\left (J^2 - 1 \right)$ \cite{kattel2024kondo,tang2024quantum}. This complex mode with finite non-zero energy describes an exponentially boundary-localized bound mode.

\begin{figure}
    \centering
\includegraphics[width=\linewidth]{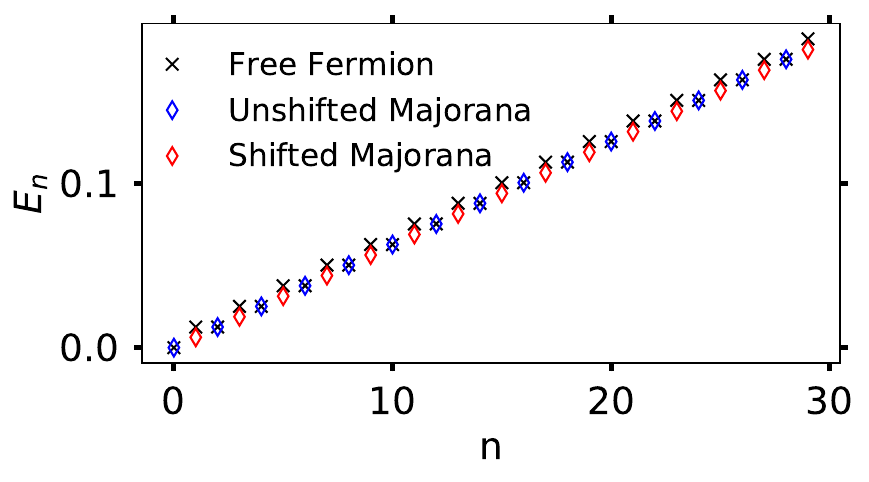}
    \caption{{\bf Excitation Spectrum}: $n$ low-lying excited states of the model in the UV fixed point ($J\to 0^+$) is the free Fermion spectrum labeled by the black cross, and the low-lying excited states in the IR ($J=1$) consists of one free Majorana chain labeled by blue diamond and one Majorana chain with phase shift labeled by red diamond.}
    \label{fig:Spectrum}
\end{figure}

Combining one phase-shifted Majorana fermion with an unshifted fermion at each site, the resulting composite object will no longer exhibit a free fermion spectrum. Instead, the interaction between the shifted and unshifted Majoranas introduces correlations that deviate from the usual Fermi liquid behavior, indicating the onset of non-Fermi liquid dynamics. The complete spectrum of the Hamiltonian in Eq.\eqref{MainH} is 
\begin{equation}
    E = \sum_{{j},a} 2 n_{{j},a}\cos k_{j,a}+E_0,
\end{equation}
where $n_{j,a} \in \{1,0\}$ depends on whether the corresponding mode is occupied or not, and $a\in\{1,2\}$ represents the two decoupled chains. The ground state energy is given by $E_0=-\frac {2 J^2\sec^{-1} (J)} {\pi\sqrt {J^2 - 1}} - \frac {4 (N + 
      1)} {\pi} + 2,$
and the low-lying single particle excitations with positive energies are shown in Fig.~\ref{fig:Spectrum} in the UV, which is obtained by taking $J\to 0^+$ where the impurity is decoupled and in the IR, which is obtained by setting $J=1$ where all the couplings in the Majorana chain depicted in Fig.~\ref{fig:pictorialHam} are the same.  When $J > \sqrt{2}$, a complex boundary mode appears with energy $E_B = \frac{J^2}{\sqrt{J^2 - 1}}$, exceeding all other modes, and wavefunction $\psi(x) \sim e^{-x \ln(\sqrt{J^2 - 1})}$ that is exponentially localized at the edge~\cite{kattel2024kondo}.

% Before computing the physical quantities showing the non-Fermi liquid behavior, let us briefly revisit the model with an additional $J \sigma^y_{0}\sigma^y_1$ coupling \textit{i.e.}
% \begin{equation}
%     H = \sum_{i=1}^{N-1} (\sigma^x_i\sigma^x_{i+1} + \sigma^y_i\sigma^y_{i+1}) +J\sigma^x_{0}\sigma^x_1+J \sigma^y_{0}\sigma^y_1
%     \label{eqn:ham2}
% \end{equation}
% studied in \cite{kattel2024kondo}. This model is shown in the pictorial description outlined above in Fig.~\ref{fig:pictorialHam} b).
% In this case, since both chains have impurities at their edges, the two Majorana fermions at each bulk site experience identical phase shifts. As a result,  these two identically shifted Majoranas can be combined to form a complex Fermion at each site, leading to a Fermi liquid behavior, as we have explicitly demonstrated in \cite{kattel2024kondo}.

We now turn to the explicit computation of physical quantities associated with the impurity to demonstrate the model's two-channel Kondo behavior. While the impact of a local magnetic field on the impurity has been explored in ~\cite{zvyagin2013possibility}, our focus here will be on impurity contribution to thermodynamic entropy.
The free energy $F$ at finite temperature $T$ for the model can be written as
\begin{equation}
    F(J,T,N) = -T\sum_{j,a}f_T(k_{j,a})
\end{equation}
with $f_T(k) = \ln[1+\exp({-\frac{2}{T}\cos k })]$ being the free energy of each mode, and quasi-momenta $k_{j,a}$ from Eq.\eqref{k1} and \eqref{quantcond}.
The impurity free energy can then be obtained by subtracting the bulk free energy $F_{\mathrm{imp}}(J,T,N) = F(J,T,N)-F(0,T,N-1)$, which can be written in the limit $N\to\infty$ as

\begin{equation}
\label{Fimp}
\begin{split}
    F_{\mathrm{imp}}(J,T) &=-T\ln 2 + \frac{T}{\pi}\int \delta(k)\frac{d}{dk}f_T(k)\mathrm{d}k,
    %\\& = -\frac{T}{2}\ln 2 - \frac{T}{\pi}\int f_T(k)\delta'(k)\mathrm{d}k
\end{split}
\end{equation}
with $\delta(k)=-\frac{i}{2}\ln A_B(k)=-\frac{i}{2}\ln \frac{(1-J^2)+e^{-2ik}}{1+(1-J^2)e^{-2ik}}$ being the phase shift which is the log of the scattering amplitude $A_B$.

Recall for the standard Kondo model, the phase shift takes the form $\delta_{\rm Kondo}(p)=\frac{\pi}{2}-\frac{p}{2T_K}$, with $T_K$ being the Kondo Temperature\cite{andrei1983solution}. 
In our case, the low energy excitations is at $k=\frac{\pi}{2}+p$ for slow momenta $|p|\ll\pi$, $\delta(p)=\frac{\pi }{2}-\frac{\left(2-J^2\right)}{J^2}p$ which shows that the phase shift is $\frac{\pi}{2}$ at low energy and the Kondo temperature $T_K=\frac{J^2}{2-J^2}$. Noticing that the Kondo temperature $T_K$ diverges at the boundary transition point $T_K(J=\sqrt{2})=\infty$ and when there is a boundary bound mode $T_K(J>\sqrt{2})<0$.

%Rewriting the integral, we obtain
% \begin{align}
%     F_{\mathrm{imp}}&= -T \ln (2)-\int_0^{2\beta}\frac{i \ln \left(\frac{1+\left(J^2-1\right) e^{-2 i \csc ^{-1}\left(\frac{2 \beta }{u}\right)}}{-J^2+e^{2 i \sec ^{-1}\left(\frac{2 \beta }{u}\right)}+1}\right)}{2 \pi  \beta( 1+ e^u)}\mathrm{d}u,
%     \label{fimprw}
% \end{align}
The low-temperature expansion for the free energy, after evaluating the integral in Eq.\eqref{Fimp}, becomes (when $J<\sqrt{2}$)
\begin{equation}
  F_{\mathrm{imp}}(J,T\to 0)= -\frac{T}{2}\ln2 
-\frac{\pi}{24\,T_K}\,T^2 + \mathcal{O}(T^3),
  \label{nnfimp}
\end{equation}
 The value at $T=0$ is 
    $F_{\mathrm{imp}}(T\to 0)= -T \ln \sqrt{2}$.
At high-temperature, the integrand $Tf'_{T}(k)\to 0$ in Eq.\eqref{Fimp} as $T\to\infty$, thus
$   F_{\mathrm{imp}}(T\to \infty)= -T \ln {2}$.

\begin{figure}
    \centering
    \includegraphics[width=1.05\linewidth]{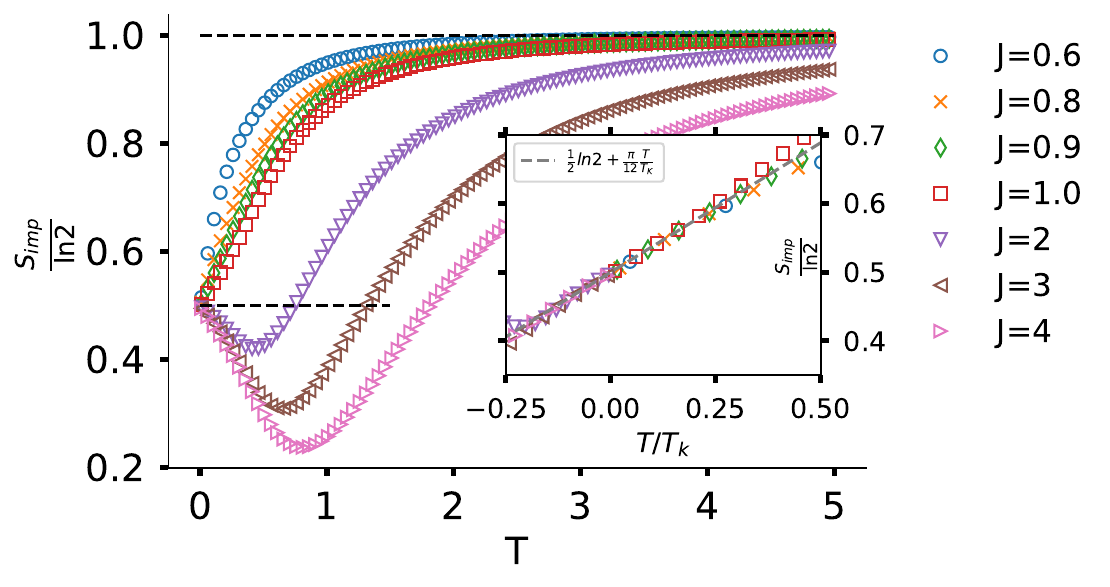}
    \caption{ {\bf{Impurity thermal entropy as a function of temperature}}. Notice that the thermal entropy of the impurity is a monotonic function of temperature when $J<\sqrt{2}$ as required by the $g$-theorem, whereas when $J>\sqrt{2}$, the entropy is not monotonous. The inset shows the universal behavior when we collapse the thermal entropy $S(T/T_K)=\frac{1}{2}\ln 2+\frac{\pi}{12}T+\mathcal{O}(T^2)$ as shown in Eq.\eqref{nnfimp}. The decrease in impurity thermal entropy when $J>\sqrt{2}$ is due to a negative Kondo temperature.}
    \label{fig:entropy}
\end{figure}

These two limiting cases show that the impurity entropy $S_{\mathrm{imp}}(T)=-\partial_T F_{\mathrm{imp}}(T)$ is $\ln\sqrt{2}$ in the IR, whereas it is $\ln 2$ in the UV. At low temperature, the impurity entropy can be expressed as a universal function in the form $S_{\mathrm{imp}}(\frac{T}{T_K})$ as shown in the inset of Fig.\ref{fig:entropy} and the full temperature-dependent impurity entropy for various values of the boundary couplings in Fig.\ref{fig:entropy} for both Kondo regime $J<\sqrt{2}$ and bound-mode regime $J>\sqrt{2}$.
The impurity entropy in the ultraviolet limit (high temperature) is $S_{\mathrm{imp}}(T\to\infty) = \ln 2$, which shows that the impurity is asymptotically free. In the infrared limit (low temperature), however, there is residual entropy $S_{\mathrm{imp}}(T\to0) = \ln \sqrt{2}$, which suggests non-Fermi liquid behavior as seen in other systems with 2-channel behavior \cite{kattel2024overscreened,schlottmann1993multichannel,granath1996two,PhysRevLett.128.146803}. Notably, this behavior occurs irrespective of the boundary couplings. For $J < \sqrt{2}$, the impurity entropy—and hence the $g$-function—monotonically decreases as the boundary renormalization group (RG) flows from the UV to IR regime, consistent with the $g$-theorem \cite{friedan2004boundary}.

The $g$-function is a measure of the degrees of freedom associated with a boundary. It appears as a volume-independent contribution to the free energy in a two-dimensional quantum field theory with boundaries~\cite{affleck1991universal,casini2016g,harper2024g,pozsgay2010mathcal}. It decreases monotonically along the renormalization group (RG) flow, reflecting the reduction in boundary degrees of freedom.  For $J > \sqrt{2}$, however, the non-monotonic behavior of $S_{\mathrm{imp}}(T)$ suggests a $g$-theorem violation. However, this can instead be attributed to a massive bound mode in the spectrum, which violates conformal symmetry, making the $g$-theorem no longer applicable. This analysis indicates that a bCFT description is valid in the $J < \sqrt{2}$ regime, while in the $J > \sqrt{2}$ regime, boundary CFT fails due to the appearance of the massive bound mode \cite{kattel2024kondo}.

We now proceed to compute the $g$-function  for $J<\sqrt{2}$ using CFT methods \cite{he2014quantum}. As shown in Fig.~\ref{fig:pictorialHam}, our system is equivalent to one decoupled Majorana fermion and two Ising CFTs, one with Neumann boundary conditions on both sides and one with a relevant perturbation at one end. The relevant perturbation due to the impurity triggers a boundary RG flow from Neumann to Dirichlet boundary conditions. To understand the effect of this boundary RG flow, let us recall that the Ising CFT has three primary fields: the identity operator \( \mathbb{I} \) with conformal dimension \( h = 0 \), the energy density operator \( \epsilon \) with \( h = 1/2 \), and the spin operator \( \sigma \) with \( h = 1/16 \) \cite{francesco2012conformal}. Moreover, for any rational CFT, the conformal boundary states satisfy~\cite{cardy2004boundary}
    $(L_n-\bar L_{-n})\ket{B}=0$,
where $L_n$ and  $\bar L_{-n}$ are the generators of the local holomorphic and antiholormophic conformal transformation. The solutions to this equation form a vector space, with the so-called Ishibashi states serving as its basis, given as ~\cite{ishibashi1989boundary}
\begin{equation}
|j\rangle\rangle \equiv \sum_{M} |j; M\rangle \otimes U |j; M\rangle.
\label{eqn:ISb}
\end{equation}
Here, $j$ labels the conformal towers, $M$ denotes the level within the conformal tower, and $U$ is an antiunitary operator formed by the combination of time reversal and complex conjugation. A true boundary state can be expressed as a linear combination of Ishibashi states Eq.\eqref{eqn:ISb} given by the Cardy condition \cite{cardy2004boundary}
$  \ket{B_\alpha}= \sum_j B_\alpha^j |j\rangle\rangle$,
with the reflection coefficient, $B_\alpha^j$  satisfying the Cardy condition. They are given by the elements of the modular S-matrix as
    $B_\alpha^j = \frac{S_{\alpha j}}{\sqrt{S_{\mathbb{I}j}}}$.
with the modular \( S \)-matrix for the Ising CFT is given by \cite{caputa2017quantum,cardy1989boundary}
\[
S =  \begin{pmatrix}
S_{\mathbb{I}, \mathbb{I}} & S_{\mathbb{I}, \epsilon} & S_{\mathbb{I}, \sigma} \\
S_{\epsilon, \mathbb{I}} & S_{\epsilon, \epsilon} & S_{\epsilon, \sigma} \\
S_{\sigma, \mathbb{I}} & S_{\sigma, \epsilon} & S_{\sigma, \sigma}
\end{pmatrix} = \frac{1}{2} \begin{pmatrix}
1 & 1 & \sqrt{2} \\
1 & 1 & -\sqrt{2} \\
\sqrt{2} & -\sqrt{2} & 0
\end{pmatrix}.
\]

The \( g \)-function characterizing the flow from the Neumann to the Dirichlet boundary conditions can directly be read off the modular S-matrix  as \cite{affleck1991universal}
\begin{equation}
g=\frac{g_{UV}}{g_{IR}}=\frac{S_{\mathbb{I}, \sigma}}{S_{\mathbb{I}, \mathbb{I}}} = \frac{\sqrt{2}/2}{1/2} = \sqrt{2}.
\end{equation}
This ratio reflects the additional boundary entropy introduced by the change from Neumann to Dirichlet boundary conditions. Notice that this change in boundary conditions occurs only in one of the two chains. Hence, the total entropy change due to the impurity is $S_{imp}=\ln g=\ln \sqrt{2}$, as previously computed using thermodynamical considerations.
% \JP{JP: I see lots of equations that are numbered and never referenced again, making it unclear why it deserves its own line.}

This prediction of the impurity entropy from bCFT can be verified directly in the lattice. To do so, we compute the difference in the Von Neumann entanglement entropy of the chain given by Hamiltonian Eq.\eqref{modelham} for impurity coupling $J=1$ and $J\to 0^+$. Notice that at the left end of the two chains, the boundary conditions are the same (Neumann), and hence, the difference in entropy is expected to vanish, while at the right end of the chain, the boundary conditions in the chain with vanishing impurity coupling are Neumann, but the one with non-vanishing impurity coupling flows to Dirichlet such that the difference in the entropy at the right end should approach the expected value of $\ln\sqrt{2}$. We compute the entanglement entropy using the density matrix renormalization group method using the ITensor library \cite{fishman2022itensor,fishman2022itensor1}. More concretely, we define
\begin{equation}
    S_{\mathrm{diff}}=S(j,J=1,N)-S(j,J\to 0^+,N)
    \label{sdiffeqn}
\end{equation}
where $S(j,J,N)$ represents the entanglement entropy of the chain governed by the spin chain Hamiltonian Eq.\eqref{modelham} with boundary coupling $J$ and the total number of sites $N$ by bi-partitioning it at every bond between sites $j$ and $j+1$. As shown in \cite{alkurtass2016entanglement,affleck2009entanglement}, the difference in the entanglement is equal to the difference of the logarithm of the g-functions in the UV and IR \textit{i.e.} $S_{\mathrm{diff}}=\ln\left(\frac{g_{UV}}{g_{IR}} \right)$. Thus, we expect this quantity to be $\ln\sqrt{2}$ near the impurity site and to vanish as $j\to N$ in the thermodynamic limit. Since there is no analytic result for boundary flow from Neumann-Neumann boundary conditions to Dirichlet-Neumann boundary conditions induced by a single impurity at the left end of the chain, let us momentarily consider the chain with two impurities at the two ends of the chain (at site $j=0$ and $j=N+1$) by considering a Hamiltonian of the form $H_{2\mathrm{imp}}=H_s+J \sigma_{N}^x\sigma^x_{N+1}$ with $H_s$ given by Eq.\eqref{modelham}. Then, the difference in entanglement entropy $\tilde {S}_{\mathrm{diff}}=\tilde {S}(j,J=1,N)- S\tilde (j,J\to 0^+, N)$ where $\tilde {S}(j,J,N)$ represents the entanglement entropy of the chain given by Hamiltonian $H_{2\mathrm{imp}}$ is expected to be $\tilde {S}_{diff}=\ln(\sqrt{2})$ independent of the site $j$ in the thermodynamic limit \cite{affleck2009entanglement} as the two impurity drive the boundary flow from Neumann-Neumann to Dirichlet-Dirichlet boundary conditions. 

\begin{figure}
    \centering
    \includegraphics[width=1\linewidth]{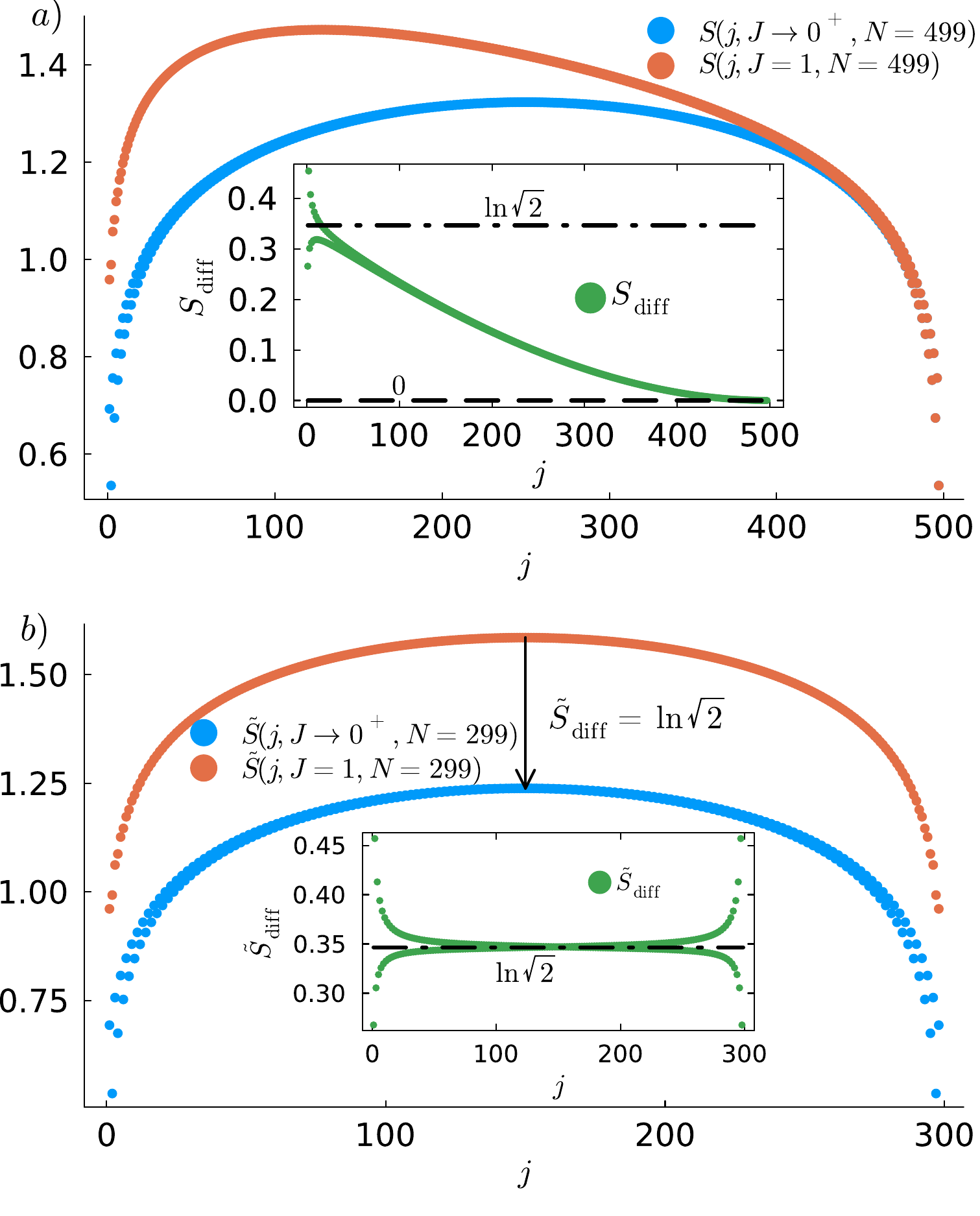}
    \caption{ {\bf{Entanglement entropy for weak and strong impurity couplings.}} a) Entanglement entropy for weak ($S(j,J\to 0^+,N)$) and strong ($S(j,J=1,N)$) couplings and the difference between them $S_{\mathrm{diff}}$ for Eq.\eqref{modelham} for $N=499$ bulk sites obtained by setting the truncation cut-off at $10^{-10}$ and performing 25 sweeps using ITensor library. Apart from the oscillation at the boundary, $S_{\mathrm{diff}}$ yields the expected asymptotic behavior at the two extreme edges of the chain: $\ln\sqrt{2}$ at the left edge with impurity and $0$ at the right edge. b) Entanglement entropy for weak ($\tilde{S}(j,J\to 0^+,N)$) and strong ($\tilde{S}(j,J=1,N)$) couplings and the difference between them $\tilde{S}_{\mathrm{diff}}$ for Hamiltonian $H_{2\mathrm{imp}}$ with $N=299$ bulk sites obtained by setting the truncation cut-off at $10^{-10}$ and performing 25 sweeps. Apart from the well-known oscillation at the boundary~\cite{laflorencie2006boundary} due to open boundary conditions, $S_{\mathrm{diff}}$ shows the expected value of $\ln\sqrt{2}$ for every site.  }
    \label{fig:entdiff}
\end{figure}

The von Neumann entanglement entropy quantifies entanglement between regions \( A \) and \( B \) in a spin chain. Given the reduced density matrix \( \rho_A = \mathrm{Tr}_B \rho \) for the ground state \( \rho = |\Psi_0\rangle \langle \Psi_0| \), the entropy is defined as
$S_{\text{vN}} = -\mathrm{Tr}(\rho_A \ln \rho_A)$. 
The relative entropy difference, \( S_{\mathrm{diff}} \), defined in Eq.~\eqref{sdiffeqn}, is computed for the Hamiltonian expressed in spin variables (Eq.~\eqref{modelham}) with a single impurity, as well as for the two-impurity Hamiltonian, \( H_{2\mathrm{imp}} \), using DMRG implemented in the ITensor library \cite{fishman2022itensor}. The results for \( S_{\mathrm{diff}} \) with \( N=499 \) and \( \tilde{S}_{\mathrm{diff}} \) with \( N=299 \) are presented in Fig.~\ref{fig:entdiff}.

The difference between the entanglement entropy between the chains at the boundary containing the impurity and those without it shows distinct behavior. Near the boundary, there is an oscillating part, but beyond that, the difference approaches \( \ln\sqrt{2} \) for the chain with the impurity, while the difference at the boundary without the impurity approaches zero, as observed for the spin Hamiltonian in Eq.~\eqref{modelham}. Similarly, for the two-impurity Hamiltonian \( H_{2\mathrm{imp}} \), the relative entropy difference approaches \( \ln\sqrt{2} \) uniformly in the bulk, with oscillations only at the boundary. This boundary oscillation is a well-known feature in systems with open boundary conditions \cite{laflorencie2006boundary}. This suggests that for both cases, the uniform part of the relative entropy difference (obtained by subtracting the oscillation due to open boundary) shows the expected behavior; thus, these results confirm that the impurity-induced \( g \)-function is \( \sqrt{2} \), consistent with predictions from boundary CFT and independently from thermodynamic consideration discussed earlier.

In conclusion, we showed that the quantum $XX$ chain with a boundary defect exhibits the critical properties of the two-channel Kondo problem for boundary coupling $J<\sqrt{2}$ where two channels of Majorana fermion chain over screens one localized Majorana fermion at the boundary as shown in Fig.~\ref{fig:pictorialHam}. In contrast, for $J>\sqrt{2}$, there exists a massive bound mode in the spectrum that explicitly violates the scale invariance, making the $g$-function (and hence the impurity entropy) non-monotonic.

\textit{Acknowledgement:}
 We thank Anirvan Sengupta and Colin Rylands for insightful discussions and valuable suggestions. This work is partially supported by NSF Career Grant No.~DMR-1941569 (J.H.P.).
\bibliography{ref}

\end{document}